\documentclass[useAMS,usenatbib]{mn2e}
\usepackage{graphicx}
\usepackage[dvips]{color}
\usepackage{amsmath,amsthm,amssymb}
\usepackage[T1]{fontenc}
\usepackage{aecompl}

\title[Collisionless Shock Formation before Breakout]
  {Collisionless Shocks and TeV Neutrinos before Supernova Shock Breakout from an Optically Thick Wind}
\author[G. Giacinti \& A. R. Bell]
  {G.~Giacinti$^1$\thanks{E-mail: gwenael.giacinti@physics.ox.ac.uk}
  and A.~R.~Bell$^1$\\
  $^1$University of Oxford, Clarendon Laboratory, Parks Road, Oxford OX1 3PU, United Kingdom}
\date{Released 2014 Xxxxx XX}

\pagerange{\pageref{firstpage}--\pageref{lastpage}} \pubyear{2014}

\def\LaTeX{L\kern-.36em\raise.3ex\hbox{a}\kern-.15em
    T\kern-.1667em\lower.7ex\hbox{E}\kern-.125emX}

\begin{document}

\label{firstpage}

\maketitle

\begin{abstract}
During a supernova explosion, a radiation-dominated shock (RDS) travels through its progenitor. A collisionless shock (CS) is usually assumed to replace it during shock breakout (SB). We demonstrate here that for some realistic progenitors enshrouded in {\it optically thick} winds, such as possibly SN~2008D, a CS forms deep inside the wind, soon after the RDS leaves the core, and therefore significantly {\it before} SB. The RDS does not survive the transition from the core to the thick wind when the wind close to the core is not sufficiently dense to compensate for the $r^{-2}$ dilution of photons due to shock curvature. This typically happens when the shock velocity is $\lesssim 0.1 {\rm c} \, (\frac{u_{\rm w}}{10\,{\rm km/s}}) (\frac{\dot{M}}{5 \cdot 10^{-4} \, {\rm M}_\odot {\rm /yr}})^{-1} (\frac{r_\ast}{10^{13}\,{\rm cm}})$, where $u_{\rm w}$, $\dot{M}$ and $r_\ast$ are respectively the wind velocity, mass-loss rate and radius of the progenitor star. The radiative CS results in a hard spectrum of the photon flash at breakout, which would produce an X-ray flash. Cosmic ray acceleration would start before SB, for such progenitors. A fraction of secondary TeV neutrinos can reach the observer up to more than ten hours {\it before} the first photons from breakout, providing information on the invisible layers of the progenitor.
\end{abstract}

\begin{keywords}
 acceleration of particles -- plasmas -- shock waves -- cosmic rays -- supernovae: general.
\end{keywords}

\section{Introduction}
\label{introduction}

Type Ib/c and II supernovae (SNe) are generated by core collapse in massive stars. When the central engine forms, a shock wave is launched through the hydrostatic core of the progenitor. The shock is radiation-dominated (or radiation-mediated), i.e. the radiation pressure in the downstream exceeds the fluid pressure~\citep{ZeldovichRaizer}. Once the radiation-dominated shock (RDS) reaches the optically thin outer layers of the stellar core or of its wind (if optically thick), photons cannot stay confined in the immediate downstream and escape ahead of the shock. This flash of photons corresponds to shock breakout (SB)~\citep{Colgate74,Falk78,KleinChevalier78,ChevalierKlein79,Ensman:1991td,Matzner:1998mg,Blinnikov:1999eb,Calzavara:2003ap,Waxman:2007rr,Katz:2009jd,Katz:2011fz,Piro:2009xi,Nakar:2010tt,Sapir:2011ds,Sapir:2013vda}. Up until now a few of them have been observed~\citep{Campana:2006qe,Gezari:2008jb,Modjaz:2008ca,Schawinski:2008ba,Soderberg:2008uh,Ofek:2010kq}, and some X-ray flashes (XRFs) may be related to SB (e.g.~\cite{Kulkarni:1998qk,Tan:2001qw,Mazzali:2008tb,Katz:2011zx}). See~\cite{Ofek:2012rw,Ofek:2013mea} and~\cite{Murase:2013kda} for radiative signatures at and following breakout.

At SB, the RDS disappears and a collisionless shock (CS) later forms~\citep{ChevalierKlein79,Ensman:1991td,Waxman:2001kt,Chevalier:2008qz}. The Larmor radius $r_{\rm L}$ of suprathermal particles is smaller than the width of the RDS, which is $\simeq \lambda c/3u_{\rm s}$ for a shock velocity $u_{\rm s}$ and photon mean free path $\lambda$~\citep{Weaver76}. On the other hand, $r_{\rm L}$ is larger than the CS width~\citep{Bell1978a,Bell1978b} and diffusive shock acceleration becomes possible. A thorough understanding of the CS formation time is then crucial to study the onset of CR acceleration, when very high energies might be reached: $\gtrsim$ TeV~\citep{Waxman:2001kt,Katz:2011zx}, PeV~\citep{Tatischeff:2009kh,Bell:2013kq}, and maybe ultra-high energies for transrelativistic SNe~\citep{Budnik:2007yh}.

Post-main-sequence mass-loss of massive stars is sufficiently high for some SN progenitors, such as some Wolf-Rayet (WR) stars, blue and red supergiants (RSG), to end up surrounded with optically thick winds~\citep{Crowther:2006dd,Langer:2012jz}. Also, remarkable outbursts can occur before the explosion, see e.g.~\cite{Ofek:2013mea} and \cite{Svirski:2014jga}. For optically thick winds, the hydrostatic surface is not observable, which complicates our understanding of the late stages of massive star evolution~\citep{Groh:2013mma}. \cite{Katz:2011zx} demonstrated that a CS must appear during or on the time scale of SB (see also~\cite{Chevalier:2011ha,Chevalier:2012zf}, \cite{Murase:2010cu}, \cite{Svirski:2012fc} and \cite{Kashiyama:2012zn}), when the RDS reaches the optically thin layers of the wind, at an optical depth $\tau \sim c/u_{\rm s} = \beta_{\rm s}^{-1}$ from the surface~\citep{Katz:2009jd}.

We demonstrate in this paper that the formation of a collisionless shock occurs significantly {\it before} SB for some progenitors enshrouded in optically thick winds: For some realistic density profiles and shock velocities, the RDS stalls in the optically thick layers of the wind. A CS forms within the broader radiation-dominated transition soon after the RDS leaves the hydrostatic core of the progenitor, at $\tau \gg \beta_{\rm s}^{-1}$. This makes SB from some optically thick winds fundamentally different from thin winds. From a theoretical perspective, it redefines the onset of CR acceleration with respect to SB, since it can start in such cases significantly before SB. This provides a new method to constrain observationally otherwise inaccessible parameters such as the radius of the invisible hydrostatic core, see below. Also, the spectrum and energy emitted from the beginning of SB are affected by the {\it earlier} presence of a radiative collisionless shock and secondaries of $\gamma$-rays from CRs injected deep within the wind.

Supernova SN~2008D/XRF~080109 may have been an event in which a CS is formed before SB, assuming a progenitor with the parameters derived in~\cite{Svirski:2014jga}.

The paper is organized as follows. In Section~\ref{thin}, we recall a few results for progenitors surrounded with optically thin winds. In Section~\ref{thick}, we discuss conditions for the formation of a collisionless shock before breakout from optically thick winds, and confirm these with numerical simulations in Section~\ref{code}. We investigate in Section~\ref{pa} particle acceleration in thick winds, and discuss in Section~\ref{obs} observational consequences of our findings.

\section{Optically thin winds}
\label{thin}

In the following, we assume a non-relativistic shock in spherical symmetry with radius $r$, where $r=0$ corresponds to the centre of the progenitor. The hydrostatic core and the wind are assumed to be fully ionized hydrogen. Assuming more realistic compositions would not change our findings. For the temperatures we consider, between $\sim$\,eV and $\sim m_{\rm e}c^2$, the opacity $\kappa$ is dominated by Thomson scattering: $\kappa = \sigma_{\rm t}/m_{\rm p}$, where $\sigma_{\rm t}$ is the Thomson cross section.

In this Section, we first consider a progenitor surrounded with an optically {\it thin} wind. SB then starts in the outer layers of the core at $\tau \approx c/3u_{\rm s}$~\citep{KleinChevalier78,ChevalierKlein79}. For such progenitors, CS are not expected to form before SB.

In Lagrangian coordinates, the acceleration of a shell of wind is $Du/Dt = \kappa \, \mathcal{F}_{\rm rad}/c - (1/\rho) \, \partial p/\partial r$, where $\mathcal{F}_{\rm rad}$ is the photon flux, $u$ the shell velocity and $p$ the fluid pressure. The maximum velocity that can be reached by a shell, initially at $r_{\rm i}$, due to the flash of photons from breakout at $t_{\rm br}$ is $u_{\max,\gamma} = \kappa \int_{t_{\rm br}}^{\infty} \mathcal{F}_{\rm rad} dt /c < \kappa \int_{t_{\rm br}}^{\infty} \mathcal{L}dt /4 \pi c r_{\rm i}^2 \propto r_{\rm i}^{-2}$ (see also~\cite{Katz:2011zx}) where $\mathcal{L}$ denotes the SN luminosity. After beginning of SB at $r_{\rm br}$, the formation of a collisionless shock is not immediate, see~\cite{KleinChevalier78} and~\cite{Chevalier:2008qz}. The $r^{-2}$ dilution of breakout photons ensures that a shell $\mathcal{S}_1$ initially at $r_1 \geq r_{\rm br}$ will catch up supersonically a shell $\mathcal{S}_2$ initially at $r_2$, with $r_2$ (sufficiently) larger than $r_1$. Despite the wind being nearly collisionless, $\mathcal{S}_1$ is prevented from going through $\mathcal{S}_2$ by electromagnetic instabilities, which gives rise to a CS. See~\cite{Waxman:2001kt} for an estimate of their growth time.

For an optically thin circumstellar medium, we confirm numerically the formation of a CS after SB (as found by~\cite{KleinChevalier78} and~\cite{Ensman:1991td}) with our 1D-spherical radiation-hydrodynamics code, described in Section~\ref{code}. In planar geometry, the $r^{-2}$ dilution factor is not present, and no CS should form, which agrees with the findings of~\cite{Sapir:2011ds}.

\section{Collisionless shocks before breakout from optically thick winds}
\label{thick}

Let us now consider progenitors surrounded with {\it optically thick} winds. SB then starts in the wind at $\tau \sim c/u_{\rm s} = \beta_{\rm s}^{-1}$~\citep{Katz:2009jd}.

We demonstrate below the central message of this paper: For some realistic optically thick circumstellar winds and for some shock velocities, the RDS does not survive long after leaving the hydrostatic core of the progenitor, and a CS forms at $\tau \gg \beta_{\rm s}^{-1}$. At $\tau \geq \beta_{\rm s}^{-1}$, radiation still plays a non-negligible role, and numerical simulations in the next Section show that such a shock resembles a 'decaying' RDS containing a CS within its --broad-- width.

Let us take $r_{\ast}\sim 10^{13}$\,cm (resp. $\sim 10^{11}$\,cm) as orders of magnitude for radii of red supergiants (resp. Wolf-Rayets). The density profiles at $r>r_{\ast}$ in the optically thick winds are poorly known and may not be $\propto 1/r^2$. However, our results do not strongly depend on them. For the numerical simulations, we take $\rho = \dot{M} / 4 \pi u_{\rm w} r^2$ for a stellar mass loss rate $\dot{M}$ and wind velocity $u_{\rm w}$. Since $\tau = \kappa \int_r^\infty \rho dr$, $r_{\rm br} \approx \kappa \dot{M} \beta_{\rm s} / 4 \pi u_{\rm w}$. Taking plausible values for a dense wind of $\dot{M} \approx 5 \cdot 10^{-4} \, {\rm M}_\odot \, {\rm yr}^{-1}$ and $u_{\rm w} \approx 10$\,km\,s$^{-1}$ (resp. $1000$\,km\,s$^{-1}$) for RSG (resp. WR) progenitors~\citep{Crowther:2006dd}, one can reach $r_{\rm br}/r_\ast \sim 10$ for $\beta_{\rm s}=0.1$, which is compatible with some interpretations of SB observations~\citep{Campana:2006qe,Wang:2006jc,Soderberg:2008uh,Modjaz:2008ca}. One may reach $r_{\rm br}/r_\ast \gg 10$ for significantly larger $\dot{M}$, see for example~\cite{Ofek:2010kq} and \cite{Svirski:2012fc}. Let us note that the following discussion is also valid for slower shocks with $\beta_{\rm s} \approx 0.01$.


\begin{figure}
\begin{center}
\includegraphics[width=0.5\textwidth]{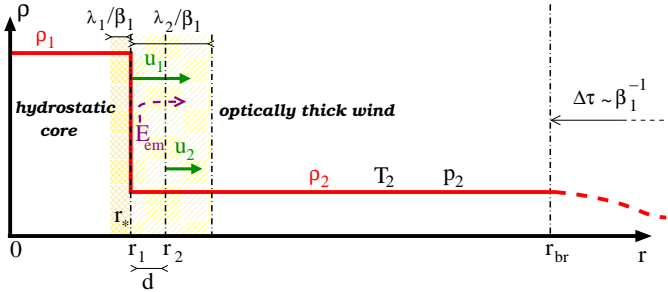}
\end{center}
\caption{Toy model of the problem, which captures the only relevant features for our analysis. The stellar core of radius $r_{\ast}=r_1(t=0)$ and density $\rho_1 = 1/\kappa\lambda_1$ is surrounded with an optically thick wind with $\rho_2 < \rho_1$. The radiation-dominated shock, with velocity $\sim u_1$ and widths $\sim \lambda_i/\beta_1$ leaves the core at $t=0$. Breakout occurs at $r_{\rm br} \gg r_{\ast}$. A collisionless shock forms when the shell at $r_1(t)$ catches up supersonically that at $r_2(t)$, which can happen at $r \ll r_{\rm br}$, see text.}
\label{Schema}
\end{figure}


In this section, we simplify the problem to elucidate the essential physics. We assume that $\rho = \rho_1$ for $r < r_\ast$ (core). Since acceleration of wind shells depends only on the flux of radiation and not on the fluid density, we take for heuristic purposes $\rho = \rho_2$ for $r_\ast \leq r \leq r_{\rm br}$ (thick wind) with $r_{\rm br} \gg r_\ast$, see Fig.~\ref{Schema}. At $t=0$, the RDS leaves the core and enters the wind. Its width becomes $\sim \lambda_2 c/u_1 =\lambda_2 / \beta_1$, where $\lambda_{1,2} = 1/\kappa\rho_{1,2}$ is the photon mean free path and $u_1$ the velocity reached by the shock after accelerating in the steep density gradient at the edge of the core --see e.g.~\cite{Sakurai60}, \cite{Matzner:1998mg}, \cite{Sapir:2011ds}, \cite{Ro:2013lfa}. In the thick wind, a CS appears if two shells catch up each other supersonically, i.e. the shell at lower radius rams into the other one with a relative velocity (much) larger than the sound speed $c_{\rm s}$. The shock is collisionless rather than viscous, see calculations in~\cite{Waxman:2001kt}. Micro-instabilities, such as the Weibel instability, mediate the formation of a CS. While this does not occur in a regular RDS, it can happen if fluid shells in the upstream cannot be accelerated to sufficiently large velocities by radiation to prevent this.

Let us consider two shells $\mathcal{S}_1$ and $\mathcal{S}_2$ with initial radii $r_1(t=0) = r_\ast$ and $r_2(t=0) = r_\ast + d$ with $d<\lambda_2 / \beta_1$, see their initial locations in Fig.~\ref{Schema}. With the above values, the mass swept up by the shock in the wind at $r \leq r_{\rm br}$ is negligible compared to the ejecta mass. $\mathcal{S}_1$ then does not significantly slow down before SB and we can assume its velocity $u_1$ to remain constant. Let us denote by $u_2$ the maximum velocity to which $\mathcal{S}_2$ can be accelerated by radiation {\it only}. In practice, $c_{\rm s} \ll u_1$ because $c_{\rm s} \approx 2 \cdot 10^5\,{\rm m\;s^{-1}}\sqrt{T_2/100\,{\rm eV}}$ in the wind heated to $T_2$. $\mathcal{S}_1$ will then catch up $\mathcal{S}_2$ and {\it create a CS before SB} if (i) $u_2<u_1 - c_{\rm s} \simeq u_1$ and (ii) the radius by which they catch up is smaller than $r_{\rm br}$: $r_\ast + \frac{d}{1-u_2/u_1} < r_{\rm br}$. In the limiting case where no absorption of radiation occurs between $\mathcal{S}_1$ and $\mathcal{S}_2$, all photons that have accelerated $\mathcal{S}_1$ may accelerate $\mathcal{S}_2$. The velocity of a fluid shell is proportional to the integrated flux of radiation passing through it. Therefore, the maximum velocity reachable by $\mathcal{S}_2$ due to these photons does not exceed
\begin{equation}
  u_{\rm 2,sph} \leq u_1 \left( \frac{r_\ast}{r_\ast + d} \right)^2\;.
\label{Eq_u2_1_over_r2}
\end{equation}
The $r^{-2}$ dilution of radiation intensity is the main reason why $\mathcal{S}_1$ may catch up $\mathcal{S}_2$. In practice, the actual velocity reached by $\mathcal{S}_2$ is larger or smaller than $u_{\rm 2,sph}$, depending on additional competing effects. First, photons pushing $\mathcal{S}_1$ lose energy, and $u_{\rm 2,sph}$ is likely to be overestimated in~(\ref{Eq_u2_1_over_r2}). Second, $\lambda_1$ increases in the expanding shocked core and additional radiation may leak out of it and accelerate both $\mathcal{S}_1$ and $\mathcal{S}_2$. Third, the dense wind between the two shells may radiate through $\mathcal{S}_2$ part of its energy $E_{\rm em}$ while being compressed, which further accelerates $\mathcal{S}_2$. The two first effects work in favour of $\mathcal{S}_1$ catching up $\mathcal{S}_2$, while the third one has the opposite effect. Therefore, we estimate the latter one, so as to know if and when a collisionless shock can form. Since $E_{\rm em} \simeq \int_{r_{\ast}}^{r_{\ast}+d} 4\pi r^2 \, \frac{\rho_2}{2} u_1^{2} \, dr$, an upper limit on the velocity of $\mathcal{S}_2$ is
\begin{equation}
u_{2} \leq u_{1} \left( \frac{r_\ast}{r_\ast + d} \right)^2 + \frac{\kappa}{c}\,\frac{E_{\rm em}}{4\pi (r_\ast + d)^2}\;.
\end{equation}
This yields
\begin{displaymath}
u_{2} \leq u_{1} \left( \frac{1}{1 + \tilde{d}} \right)^2 \left[ 1 + \frac{\beta_1}{2 \tilde{\lambda_2}} \left( \tilde{d} + \tilde{d}^2 + \frac{\tilde{d}^3}{3} \right) \right]\,,
\end{displaymath}
where we have written lengths $x$ in units of $r_\ast$ as $\tilde{x} = x/r_\ast$. This estimate is likely to overestimate $u_2$ because the surface ($\propto r_2(t)^2$) of $\mathcal{S}_2$ non-negligibly increases during its acceleration, and because not all kinetic energy will be radiated in reality. We perform numerical simulations in the next section for more accurate results. Radiation diffusing from the expanding core cannot prevent $\mathcal{S}_1$ from catching up $\mathcal{S}_2$, because $r_2(t) \geq r_1(t)$. However, a large extra acceleration might in principle make them meet beyond $r_{\rm br}$ and prevent a CS from forming at $r < r_{\rm br}$. We found this to have a negligible effect for relevant situations with our simulations.

For an optically {\it thick} wind with density proportional to $ r^{-2}$, $\tilde{d} < \tilde{\lambda_2}/\beta_1 \ll 1$ is satisfied close to the hydrostatic core. Condition (i) then becomes $\beta_1 < 4 \tilde{\lambda_2}$. Condition (ii) gives : $\beta_1 < 4 \tilde{\lambda_2} [ 1 - 1/2(\tilde{r_{\rm br}} - 1) ]$, which is not significantly more stringent than (i) for $\tilde{r_{\rm br}} \approx 10$. In practice, the main uncertainty lies in the factor '4', which is likely to be larger than this conservative estimate. From our numerical simulations, we find that approximately
\begin{equation}
\beta_1 \lesssim 10 \tilde{\lambda_2} = 0.1 \left( \frac{u_{\rm w}}{10\,{\rm km/s}} \right) \left( \frac{r_\ast}{10^{13}\,{\rm cm}} \right) \left( \frac{\dot{M}}{5 \cdot 10^{-4} \, {\rm M}_\odot {\rm /yr}} \right)^{-1}
\label{ConditionCS}
\end{equation}
for a $r^{-2}$ wind (or equivalently, $\lambda_2 / \beta_1 \gtrsim r_{\ast} / 10$), is a good overall estimate which does not noticeably depend either on the density profile of the progenitor, or on the sharpness of the transition between the core and the wind. $\beta_1$ corresponds to the velocity of the shock when it enters the wind. For simplicity, we took in this analytical discussion a flat profile for the stellar envelope, but we verified numerically that our results hold for more realistic density profiles, such as $\rho \propto r^{-2}$ for a red supergiant envelope.

Therefore, if the shock velocity does not exceed the value given by Eq.~(\ref{ConditionCS}), the RDS does not survive the transition from the core to the wind, because $\mathcal{S}_1$ catches up $\mathcal{S}_2$ at $r<r_{\rm br}$. A CS then forms in front of the expanding core. In other words, for a given wind density, and below a given shock velocity, the kinetic energy between $\mathcal{S}_1$ and $\mathcal{S}_2$ that can be radiated through $\mathcal{S}_2$ is not sufficiently large to compensate for the dilution of photons due to the spherical geometry of the problem. For progenitors with the above parameters, this yields $\beta_1 \lesssim 0.1$, which is typical of RSG shock velocities. For such progenitors, a CS forms before SB. For some WRs, larger shock velocities can occur and a CS forms before SB only for larger values of $\tilde{\lambda_2}(r=r_{\ast})$: For example, the mildly relativistic shock of SN~2006aj does not satisfy Eq.~(\ref{ConditionCS}) with $\dot{M} \approx 3 \cdot 10^{-4} \, {\rm M}_\odot \, {\rm yr}^{-1}$.

More precisely, the scenario discussed here applies when the circumstellar wind is sufficiently dense to be optically thick, but not so thick that Inequality~(\ref{ConditionCS}) is violated. For progenitors with steady winds (density $\rho \propto r^{-2}$), this typically corresponds to moderately thick winds with e.g. $r_{\rm br} \approx 10 \, r_{\ast}$. Let us mention that in the hypothetical case of a progenitor surrounded by a thick wind with a density profile flatter than $r^{-2}$ at $r<r_{\rm br}$ (e.g. due to variations in time of $\dot{M}$ before the explosion), this scenario can also be valid for significantly larger values of $r_{\rm br}/r_{\ast}$: It is valid as long as the condition $\beta_1 \lesssim 10 \tilde{\lambda_2}$ is satisfied close to the core.

Red supergiants or Wolf-Rayet stars with relatively high mass-loss rates prior to the explosion are good candidates : Moderately thick winds with $r_{\rm br} \approx 10 \, r_{\ast}$ correspond to $\dot{M}$ in the range from a few~$\times 10^{-5} \, {\rm M}_\odot \, {\rm yr}^{-1}$ to $\approx 10^{-3} \, {\rm M}_\odot \, {\rm yr}^{-1}$ for typical values of $r_{\ast}$, $u_{\rm w}$ and $\beta_1$ for RSGs and WRs. For example, Type~Ibc supernova SN~2008D/XRF~080109 may have been an event in which a CS is formed before SB : \cite{Svirski:2014jga} suggest that SN~2008D is consistent with the explosion of a WR in a moderately thick wind, and that the progenitor underwent a steady but enhanced mass-loss during the last $\lesssim 10$ days. Very interestingly, the parameters derived by~\cite{Svirski:2014jga} for SN~2008D ($\beta_1 \approx 0.25$, $u_{\rm w} \approx 1000$\,km\,s$^{-1}$, and $\dot{M} \approx 2 \cdot 10^{-4} \, {\rm M}_\odot \, {\rm yr}^{-1}$ close to the star) marginally satisfy Inequality~(\ref{ConditionCS}) for $r_{\ast} \approx 10^{11}$\,cm. This gives an additional and important reason to search for SN~2008D-like events in the future. The rate of such events depends on the likelyhood for a star to undergo enhanced mass-loss during the last few weeks or years preceding the supernova.

In contrast, progenitors of Type~IIn supernovae are not expected to satisfy our scenario, since Eq.~(\ref{ConditionCS}), with typical values of $\dot{M}$ and $u_{\rm w}$ for Type~IIn, implies upper limits on $\beta_1$ well below the actual shock velocities. 
For progenitors with significant mass-loss rates, the radiation-dominated shock should always survive the transition from the core to the optically thick wind, and the picture would then be a conventional one : In such cases, photons are supplied by the immediate downstream of the RDS in the thick part of the wind. For example, Type~IIn supernova SN~2010jl does not satisfy Eq.~(\ref{ConditionCS}) because of the large mass-loss rate $\dot{M} \sim 1\, {\rm M}_\odot \, {\rm yr}^{-1}$ \citep{Zhang:2012ba,Ofek:2013afa}.

In stellar cores, RDS are stable because $\lambda$ is sufficiently small to prevent conditions similar to those of Eq.~(\ref{ConditionCS}) from being met.


\begin{figure*}
\begin{center}
\includegraphics[width=0.33\textwidth]{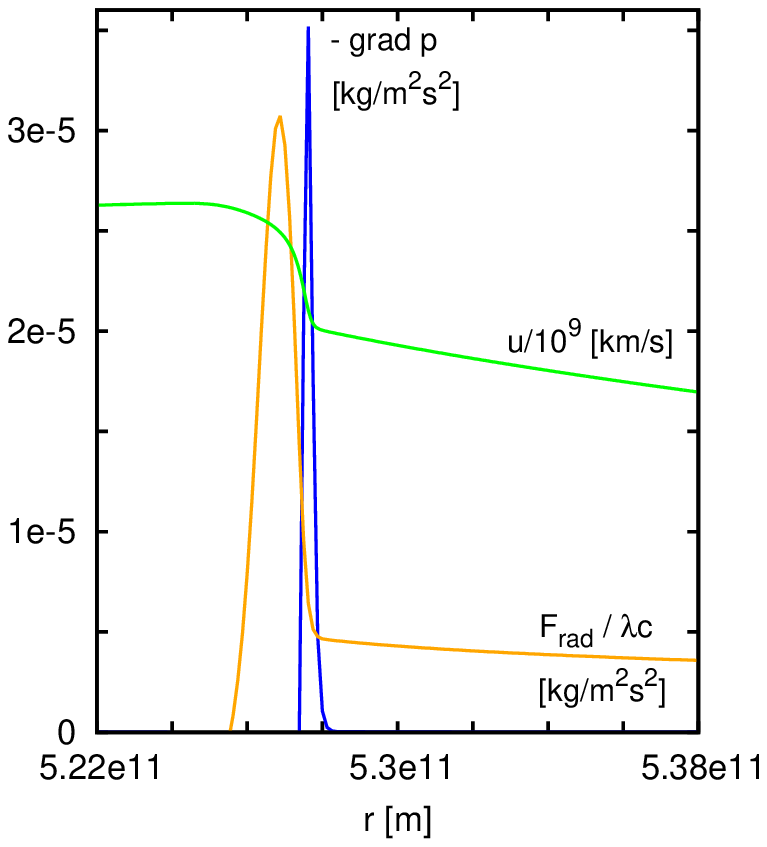}
\hskip1.4cm
\includegraphics[width=0.33\textwidth]{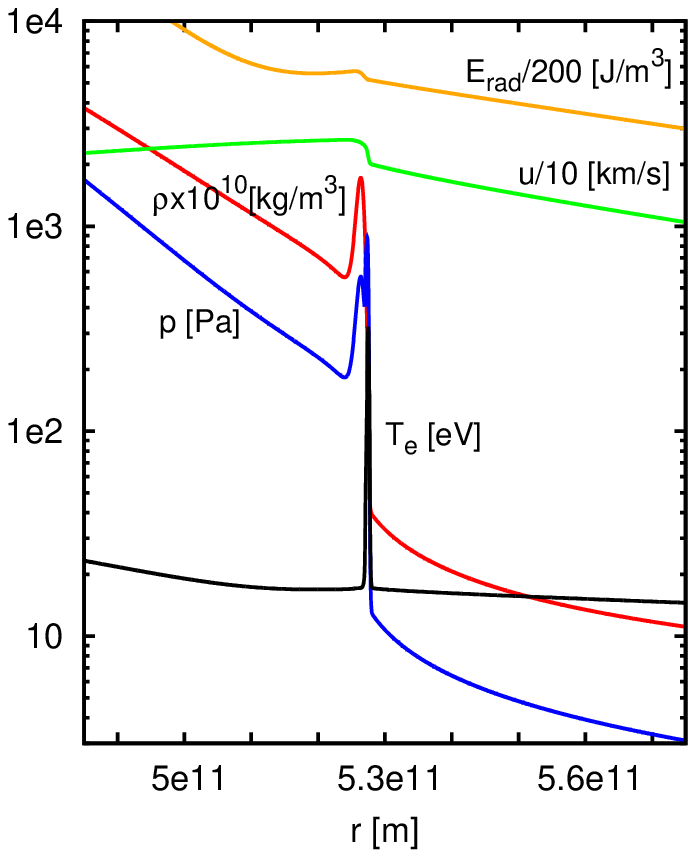}
\end{center}
\caption{Simulation of a RSG explosion in an optically thick wind {\it before} shock breakout, see text for parameters. Zoom around the downstream of the radiation-dominated transition of the shock, where a discontinuity (CS) can be seen around $r \approx 5.28\cdot 10^{11}$\,m, at $\approx 1.6 \, r_{\ast}$ and $\approx r_{\rm br}/2$. {\it Left panel:} Most of the fluid acceleration $d(\rho u)/dt$ is due to the radiation, except in a thin zone at the CS, where the fluid contribution $-\hbox{\boldmath$\nabla$}p$ dominates. It coincides with the sharp discontinuity in $u$. {\it Right panel:} A dense shell appears in $\rho$, in the downstream of the CS. Peaks in $p$ and in the electron temperature $T_{\rm e}$ appear in the CS immediate downstream. $T_{\rm rad}$ follows $T_{\rm e}$ except in the peak region, where it remains flat ---$E_{\rm rad}$ is smooth.}
\label{NumSim}
\end{figure*}


\section{Numerical simulations}
\label{code}

We confirm numerically the predictions of Section~\ref{thick} with our Eulerian 1D-spherical radiation-hydrodynamics code. The fluid is assumed to be fully ionized. The code is two-temperature, i.e. we assume proton and electron temperatures to be equal. The results presented below are not affected by this assumption. We use a gray frequency average for the radiation, and represent it by its internal energy $E_{\rm rad}$ with characteristic temperature $T_{\rm rad} = ( cE_{\rm rad} / 4\sigma )^{1/4}$, where $\sigma$ is the Stefan-Boltzmann constant. At each time step, the radiation transport is solved using a 'square-root'~\citep{Morel2000} flux-limited diffusion approximation, and the opacity $\kappa$ is assumed to be dominated by Thomson scattering. For the transfer of energy between fluid and radiation, we take into account Compton cooling and bremsstrahlung, using the formulas of~\cite{ChevalierKlein79}. Underlying assumptions for an equivalent code can be found in~\cite{ChevalierKlein79}.

Our main result does not strongly depend on the density and temperature profile of the hydrostatic core, and we take as an example those for the RSG used in~\cite{ChevalierKlein79}. We choose initial conditions such that the shock velocity reaches $\beta_{\rm s} \approx 0.1$. We use winds with density profiles $\propto r^{-2}$. We take here $\dot{M} = 5 \cdot 10^{-4}$\,M$_\odot$\,yr$^{-1}$ and $u_{\rm w}=10$\,km\,s$^{-1}$, so $\tilde{r_{\rm br}} \approx 3$. A CS appears before SB, at $\tilde{r}<1.5$. At the CS formation time, photons still have not started to escape from the optically thick material, contrary to e.g. expectations for breakout from a stellar surface. Fig.~\ref{NumSim} shows the CS at $\tilde{r} \approx 1.6$, near the downstream of the radiation-dominated transition (remains of the initial RDS). See caption of Fig.~\ref{NumSim} for details. It appears as a growing discontinuity in the smoother velocity profile. The radiation-dominated transition extends to radii larger than shown in Fig.~\ref{NumSim}. At such an early time, the CS downstream temperature is only $\sim 1$\,keV because the radiation still provides most of the fluid acceleration in its upstream, but we find the discontinuity in $u$ to grow and the CS processes a significantly larger fraction of $\rho u_{\rm s}^{2}$ at larger $r$. For other parameter values (smaller $\beta_{\rm s}$ for $\dot{M}$, $u_{\rm w}$ fixed), radiation plays a smaller role, allowing the CS to emit in hard X-rays ($\gtrsim 10$s\,keV) before its photons break out. The formula $\beta_{\rm s} \lesssim 10 \tilde{\lambda_2}$ and conclusions of the previous section have been verified by scanning the parameter ranges. We find that for significantly larger $\dot{M} = (1,\,5) \cdot 10^{-3}$\,M$_\odot$\,yr$^{-1}$ and the same core profile, the RDS survives the transition to the wind. Details of the hydrostatic core are not found to be very important, and we test the WR case by rescaling, as a first approximation, the above profile to $r_\ast=10^{11}$\,cm and larger densities. We vary $\beta_{\rm s}$ by slightly changing the explosion energy. For $\dot{M} = 5 \cdot 10^{-4}$\,M$_\odot$\,yr$^{-1}$ and $u_{\rm w}=1000$\,km\,s$^{-1}$, we find that for $\beta_{\rm s} \gtrsim 0.15$, no CS appears before SB, whereas they do appear before SB for $\beta_{\rm s} \lesssim 0.1$. Varying $\dot{M}$ and $u_{\rm w}$, also corroborates the picture announced in the previous section.

We verified numerically that once the CS is formed (even when this happens at $r < r_{\rm br}$), it survives to $r \gg r_{\rm br}$ in winds with $\rho \propto r^{-2}$ : If the shock does not sufficiently slow down at $r \leq r_{\rm br}$ and if the wind density does not increase with $r$, no process reduces the difference in velocities between the CS immediate downstream and immediate upstream down to a subsonic value. In the simulations, this difference in velocities is, on the contrary, found to grow at $r \lesssim r_{\rm br}$.

\section{Particle acceleration}
\label{pa}

Assuming conservatively a magnetic field strength at the CS similar to that at the stellar surface, $B_{\rm s} \sim 10$\,G~\citep{Barvainis1987}, and wind densities $\rho \sim 10^{-11\,(-9)}$\,g\,cm$^{-3}$, the CS is super-Alfv\'enic. Once it is formed, CR acceleration may start. Coulomb losses for suprathermal particles are sufficiently small here and do not prevent them from entering diffusive shock acceleration and being accelerated. However, for WRs with $\dot{M} \gtrsim 10^{-3}$\,M$_\odot$\,yr$^{-1}$, such losses start to inhibit CR acceleration before SB. Some findings of \cite{Waxman:2001kt} and \cite{Katz:2011zx} can be transposed to our study, yet we deal here with a shock propagating in denser regions of the wind. Assuming Bohm diffusion for CRs at the CS~\citep{oai:arXiv.org:1301.3173,Caprioli:2014tva}, and equal dwell times in the downstream and the upstream, one finds a typical acceleration time
\begin{equation}
\tau_{\rm CR} \approx \frac{8E_{\rm CR}}{3eB_{\rm s}u_{\rm s}^{2}} \approx 30\,{\rm s}\, \left( \frac{E_{\rm CR}}{10\,{\rm TeV}} \right) \left( \frac{B_{\rm s}}{10\,{\rm G}} \right)^{-1} \left( \frac{\beta_{\rm s}}{0.1} \right)^{-2}
\end{equation}
for protons. This time can be optimistic when the discontinuity in velocity at the shock is still small due to smoothing by radiation. However, magnetic field amplification at the shock due to the non-resonant hybrid (NRH) instability~\citep{Bell2004} plays a role in the opposite direction by diminishing $\tau_{\rm CR}$ and thereby facilitating CR acceleration, see~\cite{GGABprep} for a detailed study. Magnetic field amplification is (constantly) driven by the escape of the highest energy CRs in the upstream of the collisionless shock, see~\cite{Bell:2013kq}. For the ranges of parameter values that are relevant here, the typical growth time of the NRH instability is smaller than the damping time of the turbulence by the radiation field, which energy density is $U_{\rm rad} \approx \rho u_{\rm s}^{2}$. Therefore, magnetic field amplification should occur in such conditions. In the upstream of the CS, a turbulent fluid parcel with velocity $u_{\rm t}$ suffers momentum losses due to radiation (second order Fermi for photons). From the momentum equation of the fluid parcel, one can deduce the typical damping time of the turbulence :
\begin{equation}
\tau_{\rm damp} = \frac{u_{\rm t}}{du_{\rm t}/dt} \approx \frac{c^{2}}{\kappa \rho u_{\rm s}^{2}u_{\rm t}} \gtrsim \frac{c^{2}}{\kappa \rho u_{\rm s}^{3}} \; .
\end{equation}
The size of the discontinuity in velocity at the CS may be written as $\Delta u = \frac{3}{4} f \, u_{\rm s}$, where $0 < f \leq 1$ and $f = 1$ is the limiting case where no radiation accelerates the upstream of the CS. The growth rate of the fastest growing mode of the NRH instability is equal to $\gamma_{\max} = 0.5 j_{\rm CR} \sqrt{\mu_{0} / \rho}$, where $j_{\rm CR} \simeq 0.03 \, \rho f^{2} u_{\rm s}^{3} e / E_{\rm CR}$ is the CR current density which drives it~\citep{Bell:2013kq}. The instability growth time, $\tau_{\rm NRH} \approx 5 \gamma_{\max}^{-1}$, is then small compared to $\tau_{\rm damp}$ :
\begin{displaymath}
\frac{\tau_{\rm NRH}}{\tau_{\rm damp}} \lesssim \frac{10 \, \kappa \sqrt{\rho} E_{\rm CR}}{0.03 \, c^{2} e\sqrt{\mu_{0}} f^{2}} \simeq 0.08 \, \left( \frac{E_{\rm CR}}{10\,{\rm TeV}} \right) \left( \frac{f}{0.05} \right)^{-2}
\end{displaymath}
\begin{equation}
\times \left( \frac{\dot{M}}{5 \cdot 10^{-4} \, {\rm M}_\odot {\rm /yr}} \right)^{1/2} \left( \frac{u_{\rm w}}{10\,{\rm km/s}} \right)^{-1/2} \left( \frac{r}{10^{13}\,{\rm cm}} \right)^{-1}
\end{equation}
numerically for a wind with $\rho \propto r^{-2}$.

Let us note that 10\,TeV energies are reachable before breakout because $\tau_{\rm CR} \ll (r_{\rm br}-r_\ast)/u_{\rm s} \approx$ several hours (resp. minutes) for RSG (resp. WR) progenitors with $\beta_{\rm s}=0.1$ and $\tilde{r_{\rm br}} \approx 10$. For such RSGs, $\tau_{\rm CR}$(at 10\,TeV) is smaller than energy loss times from pion production through inelastic $pp$ and $p\gamma$ collisions. The typical life time of a CR proton due to $pp$ collisions, $\tau_{\rm pp} \simeq m_{\rm p}/0.2c\rho \sigma_{\rm pp}$, is
\begin{equation}
\tau_{\rm pp} \approx 4\,{\rm min}\, \left( \frac{u_{\rm w}}{10\,{\rm km/s}} \right) \left( \frac{r}{10^{13}\,{\rm cm}} \right)^{2} \left( \frac{\dot{M}}{5 \cdot 10^{-4} \, {\rm M}_\odot {\rm /yr}} \right)^{-1} \; .
\end{equation}

The background $\sim 10$\,eV photons in the thick wind are not sufficiently energetic to trigger pion production through inelastic $p\gamma$ scattering. For 10\,TeV CRs, $\gtrsim 10$\,keV photons are required to exceed the threshold for pion production. Photons with such energies can be produced by the radiative CS. However, the number density of target photons $n_{\gamma}$ must be much less than $\rho u_{\rm s}^2/h\nu$~\citep{Katz:2011zx}. We find for the typical life time of a CR proton due to $p\gamma$ collisions, $\tau_{\rm p \gamma} \simeq 1/0.2cn_{\gamma} \sigma_{\rm p \gamma}$ :
\begin{displaymath}
\tau_{\rm p \gamma} \gtrsim 2\,{\rm min}\, \left( \frac{u_{\rm w}}{10\,{\rm km/s}} \right) \left( \frac{r}{10^{13}\,{\rm cm}} \right)^{2} \left( \frac{\dot{M}}{5 \cdot 10^{-4} \, {\rm M}_\odot {\rm /yr}} \right)^{-1}
\end{displaymath}
\begin{equation}
\times \left( \frac{\beta_{\rm s}}{0.1} \right)^{-2} \left( \frac{E_{\rm CR}}{10\,{\rm TeV}} \right)^{-1} \; .
\end{equation}
$e^{\pm}$ pair creation due to $p\gamma$ interactions does not yield a stronger constraint.

In the case of Wolf-Rayet progenitors with the above parameters, $\tau_{\rm pp,p \gamma} \gtrsim 3$\,s. Consequently, TeV energies may be reached for WRs.

\section{Observational consequences}
\label{obs}

A discussion on progenitors for which the CS is expected to form before SB may be found at the end of Section~\ref{thick}, from Eq.~(\ref{ConditionCS}). We now describe the two main observational consequences : X-ray flashes and high-energy neutrinos.

For fast shocks ($\beta_{\rm s} \gtrsim 0.1 - 0.2$), RDS start to depart from thermal equilibrium, which may produce XRFs in association with SBs~\citep{Weaver76,Sapir:2011ds}. For slower shocks, UV photons are typically expected. We predict that even for lower $\beta_{\rm s}$, photons with energies $\gtrsim (1 - 10{\rm s})$\,keV can be emitted from the {\it beginning} of SB, but for a different reason: This happens when the radiative CS forms significantly before SB, such as for RSGs (or WRs with 'slow' shocks) surrounded with optically thick winds. It would heat the plasma at $r \ll r_{\rm br}$ to temperatures higher than expected for a 'slow' RDS ($\sim 10 - 100$\,eV) and load the thick wind with (hard) X-rays. This would result in a flash at breakout that both contains hard photons reflecting the presence of the hot downstream of the CS, and softer photons (notably UV) from the remains of the former RDS. The energy radiated at breakout is typically $\sim 10^{45-47}$\,erg depending on the tested progenitors --see also~\cite{Katz:2011fz} and \cite{Sapir:2013vda}. We find that, from the {\it beginning} of SB, the fraction of the energy emitted in X-rays is already roughly comparable with that in softer photons: From $\approx 10$\% to more than a half, with the largest fractions also corresponding to the highest maximum X-ray energies. The X-ray flux rises abruptly on a time scale $\sim r_{\rm br}/c \approx 30$\,s\,($\frac{r_{\rm br}}{10^{12}\,{\rm cm}}$), and then decays more slowly due to the persistence of inverse Compton on background photons in a wind with $\rho \propto r^{-2}$. Observations in different energy bands will be needed to distinguish between progenitors following the scenario presented here, and those for which the CS only starts to form {\it during} breakout, as suggested by~\cite{Katz:2011zx}. The production history of CRs at $\tau \gg \beta_{\rm s}^{-1}$ should also leave imprints in the spectrum at SB : Secondary $\gamma$-rays, notably from $\pi^{0}$ decay, are injected in the wind and partly reprocessed to lower energies through $e^{\pm}$ cascades on the large photon background ($\gamma + \gamma_{\rm b} \rightarrow e^{+} + e^{-}$).

The fact that outer layers of the thick wind at $r \lesssim r_{\rm br}$ may have been mostly accelerated by the CS implies that the energy {\it radiated} at SB may be $\ll 4\pi r_{\rm br}^{2} c^{2} \beta_{\rm s}/\kappa$. This may ease the tension between radiated energy and duration of the emission for XRF~080109 (see e.g.~\cite{Sapir:2013vda}), provided this event corresponds to SB from such an optically thick wind. The relatively low photon flux at breakout would be consistent with more energy being in the thermal plasma behind a CS, as expected in our scenario.

This work also provides a new technique to access information on SN progenitors inside thick winds, such as the radius of the stellar core $r_\ast$, and the density profile at $\tau \gtrsim \beta_{\rm s}^{-1}$. By detecting secondary $\gtrsim 100\,{\rm GeV}-1$\,TeV neutrinos (from notably $\pi^{\pm}$ decay) before the first photons from breakout, one will improve our knowledge of the still poorly understood late stages of massive star evolution. The time interval between the arrival of the first neutrinos and photons is $\Delta t_{\nu \gamma} \approx (r_{\rm br} - r_\ast) (\beta_{\rm s}^{-1}-1)/c \approx 8$\,hr (resp. 5\,min) for RSGs (resp. WRs) with the above parameters, $\tilde{r_{\rm br}} = 10$ and $\beta_{\rm s} = 0.1$. Assuming that 5\,\% of the energy processed by the shock is channelled into CRs, we typically find for a source at distance $l$, and a processed mass between $r_\ast$ and $r_{\rm br}$ of $\approx 10^{-5}$\,M$_\odot$, that $\sim 10^{3}\, (3\,{\rm kpc}/l)^{2}$ neutrinos with $\sim$\,TeV energies would be detectable {\it before SB} by IceCube or KM3NeT. One could record a few of such neutrinos for an event in the Magellanic Clouds. The low rate of such supernovae within $\simeq 100$\,kpc from Earth is the main limitation to the detection of these neutrinos with a km$^3$ observatory. For example, the rate of 2008D-like supernovae in our Galaxy should be at most 1/1000\,yr.

A supernova detected in neutrinos in the pre-shock breakout phase will generally yield more neutrinos in the post-shock breakout phase, except if the density of the progenitor wind suddenly falls sharply just beyond $r_{\rm br}$. These latter neutrinos will be detected after SB has started. For example, in a steady wind with density $\rho \propto r^{-2}$, the mass processed in the post-shock breakout phase by a shock travelling between $r_{\rm br}$ and $r > r_{\rm br}$, is $\approx \dot{M} (r - r_{\rm br})/u_{\rm w}$. This amount quickly exceeds that processed in the pre-shock breakout phase ($\sim \dot{M} (r_{\rm br}-r_\ast)/u_{\rm w}$). \cite{Murase:2010cu} studied in detail the post-shock breakout emission of neutrinos for shocks interacting with dense circumstellar material, such as shells.


\section{Conclusions}
\label{conclusions}

During a core-collapse supernova, a radiation-dominated shock propagates through the progenitor star. If the surrounding wind is optically thin, this shock stalls when it reaches the outer layers of the stellar core. In the upstream, the circumstellar material is then accelerated by escaping photons from shock breakout to a velocity roughly $\propto r^{-2}$, where $r$ is the distance to the centre of the progenitor. The outer layers of the shocked core ram supersonically into these slower layers of the wind, and a collisionless shock is expected to form during or on the time scale of supernova shock breakout. See, for example,~\cite{ChevalierKlein79}, \cite{Ensman:1991td} and \cite{Waxman:2001kt} for detailed studies.

In the present paper, we have investigated the case of supernovae occuring in thick winds. In this case, the formation of a CS should also occur no later than during or on the time scale of shock breakout ---from the 'outer' layers of the optically thick part of the wind.

We have demonstrated here that for some astrophysically-relevant progenitors surrounded with thick winds, a collisionless shock forms well before breakout, providing new ways to study invisible layers of their winds and to constrain stellar evolution theories. In such cases, the RDS has been found to stall when entering the optically thick part of the wind, notably because of shock curvature. Photons are then mostly supplied by the shock propagating in the core, and the wind is not sufficiently dense to compensate for the $r^{-2}$ dilution of photons in the wind. On the contrary, for progenitors where the RDS survives the transition from the core to the wind, such as for type IIn supernovae, photons are mostly supplied by the immediate downstream of the shock in the wind.

We have discussed, in Section~\ref{pa}, the onset of particle acceleration at the CS. For example, we predict that for some red supergiants surrounded with thick winds, a fraction of secondary high-energy neutrinos from CRs can arrive $\sim 10$ hours before photons from shock breakout, and more neutrinos are produced later in the post-shock breakout phase.

We find that the CS forms after the RDS exits the core, and before breakout, for progenitors with shock velocities $\lesssim 0.1 {\rm c} \, (\frac{u_{\rm w}}{10\,{\rm km/s}}) (\frac{\dot{M}}{5 \cdot 10^{-4} \, {\rm M}_\odot {\rm /yr}})^{-1} (\frac{r_\ast}{10^{13}\,{\rm cm}})$, where $u_{\rm w}$, $\dot{M}$ and $r_\ast$ respectively denote the wind velocity, mass-loss rate and radius of the hydrostatic core. The wind has to be sufficiently dense to be optically thick but not excessively. For progenitors with steady winds ($\rho \propto r^{-2}$), this corresponds to moderately thick winds, where e.g.\footnote{Remark : Eq.~(\ref{ConditionCS}) can also be satisfied for $r_{\rm br}/r_{\ast} \gg 10$ if e.g. the wind density profile happens to be flatter than $r^{-2}$ at $r<r_{\rm br}$.} $r_{\rm br} \approx 10 \, r_{\ast}$. Progenitors of Type~IIn supernovae are expected to have too dense winds to form CS when the RDS leave their cores. However, Wolf-Rayet stars or red supergiants with either dense winds or enhanced mass-loss prior to the explosion are better candidates. For example, Type~Ibc supernova SN~2008D/XRF~080109 has been interpreted by~\cite{Svirski:2014jga} as the explosion in a moderately thick wind of a WR star, undergoing an enhanced mass-loss during its last $\lesssim 10$ days. Interestingly, the parameters inferred by~\cite{Svirski:2014jga} for SN~2008D make it marginally consistent with the above condition. This is another important motivation to search for similar events. In the future, one can notably use them to study the formation times of collisionless shocks with respect to the photon flashes at breakout.

More generally, supernovae occurring in dense winds are promising targets for multi-messenger studies. The detection of their UVs, X-rays, $\gamma$-rays and TeV neutrinos will allow one to test a wide variety of physical and astrophysical phenomena in extreme conditions, such as particle acceleration, magnetic field amplification and shock physics.

Finally, studying CR acceleration in dense winds is important, because it should lead to a better understanding of the knee in the CR spectrum, see e.g.~\cite{Sveshnikova:2003sa}, \cite{Bell:2013kq}, \cite{Murase:2013kda}.

\section*{Acknowledgements}

We thank an anonymous referee for suggestions which improved the quality of the paper. We also thank Brian Reville and Klara Schure for useful discussions. This work was funded by the European Research Council under the European Community's Seventh Framework Programme (FP7/$2007-2013$) / ERC Grant agreement No. 247039, and supported in part by the National Science Foundation under Grant No. PHYS-1066293 and the hospitality of the Aspen Center for Physics.

\label{lastpage}

\end{document}